\documentclass[twoside,leqno,twocolumn]{article}
\usepackage{ltexpprt}

\usepackage{booktabs} 
\usepackage{algorithm} 
\usepackage{algorithmic} 
\usepackage{mdwlist} 
\usepackage[mathscr]{euscript} 
\usepackage{amsbsy} 
\usepackage{amsfonts} 
\usepackage[center]{subfigure} 
\usepackage{graphicx}
\usepackage{amsmath} 
\usepackage{amssymb}
\usepackage{xspace}
\usepackage{url}
\usepackage{threeparttable}
\usepackage{tablefootnote}
\usepackage{color}
\usepackage{tabularx}
\usepackage{multirow}
\usepackage{mathtools}
\usepackage{caption}

\usepackage{titlesec}
\titlespacing{\section}{0pt}{2pt}{1pt}
\titlespacing*{\subsection}{0pt}{2pt}{1pt}
\titlespacing*{\subsubsection}{0pt}{2pt}{1pt}

\usepackage{etoolbox}

\makeatletter
\patchcmd{\ttlh@hang}{\parindent\z@}{\parindent\z@\leavevmode}{}{}
\patchcmd{\ttlh@hang}{\noindent}{}{}{}
\makeatother

\newcommand*{\QEDA}{\small\hfill\ensuremath{\blacksquare}}%

\newcommand{\sure}{\textsc{SuRe}\xspace}

\newcommand{\model}{\textsc{Random Walk with Extended Restart}\xspace}

\newcommand{\rwer}{\textsc{RWER}\xspace}

\newcommand{\vect}[1]{\mathbf{#1}}

\newcommand{\mat}[1]{\mathbf{#1}}
\newcommand{\mati}[1]{\mathbf{#1}^{-1}}

\newcommand{\A}{\mat{A}}
\newcommand{\nA}{\tilde{\mat{A}}}
\newcommand{\MM}{\mat{M}}
\newcommand{\rr}{\mat{r}}

\newcommand{\cc}{\mat{c}}
\newcommand{\oo}{\mat{o}}
\newcommand{\qq}{\mat{q}}

\newcommand{\OUT}{\textbf{OUT}}
\newcommand{\IN}{\textbf{IN}}

\newcommand{\redtext}[1]{\textcolor{red}{#1}}

\newcommand{\hide}[1]{}

\DeclareMathOperator*{\argmin}{arg\,min}

\begin{document}

\title{\Large Supervised and Extended Restart in Random Walks \\for
	Ranking and Link Prediction in Networks}
	\author{Woojeong Jin\thanks{Seoul National University, Email: woojung211@snu.ac.kr; jinhongjung@snu.ac.kr; ukang@snu.ac.kr}
		\and
		Jinhong Jung\footnotemark[1]
		\and
		U Kang\footnotemark[1]}
\date{}

\maketitle


\fancyfoot[R]{\footnotesize{\textbf{Copyright \textcopyright\ 2018 by SIAM\\
Unauthorized reproduction of this article is prohibited}}}





\begin{abstract}
Given a real-world graph, how can we measure relevance scores for ranking and link prediction?
%
Random walk with restart (RWR) provides an excellent measure for this and has been applied to various applications such as friend recommendation, community detection, anomaly detection, etc.
However, RWR suffers from two problems:
1) using the same restart probability for all the nodes limits the expressiveness of random walk,
and 2) the restart probability needs to be manually chosen for each application without theoretical justification.

We have two main contributions in this paper.
First, we propose \model (\rwer), a random walk based measure which improves the expressiveness of random walks by using a distinct restart probability for each node.
The improved expressiveness leads to superior accuracy for ranking and link prediction.
Second, we propose \sure (\underline{Su}pervised \underline{Re}start for \rwer), an algorithm for learning the restart probabilities of \rwer from a given graph.
\sure eliminates the need to heuristically and manually select the restart parameter for RWER.
Extensive experiments show that our proposed method provides the best performance for ranking and link prediction tasks, improving the MAP (Mean Average Precision) by up to 15.8\% on the best competitor. 
\end{abstract}

\section{Introduction}
\label{sec:intro}

How can we measure effective node-to-node proximities for graph mining applications such as ranking and link prediction?
Measuring relevance (i.e., proximity or similarity) scores between nodes is a fundamental tool for many graph mining applications~\cite{adamic2003friends,agarwal2006learning,agarwal2007learning,kleinberg1999authoritative,backstrom2011supervised,AxiomSimTr}.
Among various relevance measures, Random Walk with Restart (RWR)~\cite{haveliwala2002topic,jung2016random,DBLP:conf/sigmod/JungPSK17} provides useful node-to-node relevance scores by considering global network structure~\cite{he2004manifold} and intricate edge relationships~\cite{tong2006center}.
RWR has been successfully exploited in a wide range of data mining applications such as ranking~\cite{tong2006fast,shin2015bear,jung2016random}, link prediction~\cite{adamic2003friends,agarwal2006learning,agarwal2007learning,li2016quint,backstrom2011supervised}, community detection~\cite{andersen2006local,zhu2013local}, anomaly detection~\cite{sun2005neighborhood}, etc.

However, RWR has two challenges for providing more effective relevance scores.
First, RWR assumes a fixed restart probability on all nodes, i.e., a random surfer jumps back to the query node with the same probability regardless of where the surfer is located.
This assumption prevents the surfer from considering the query node's preferences for other nodes, thereby limiting the expressiveness of random walk for measuring good relevance scores.
Second, RWR requires users to heuristically select the restart probability parameter without any theoretical guide or justification to choose it.

\begin{figure}[!t]
	
	\centering
	{\includegraphics[width=0.9\linewidth]{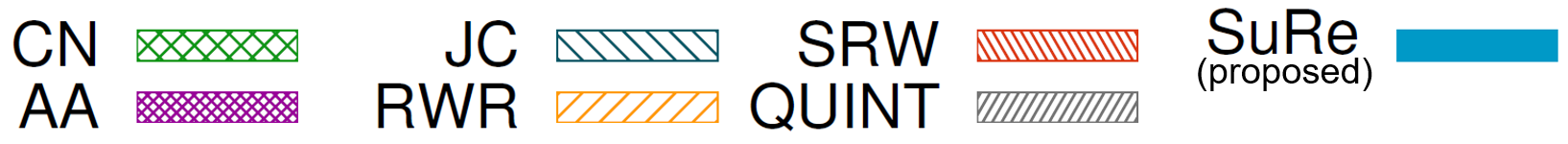}}
	\includegraphics[width=0.8\linewidth]{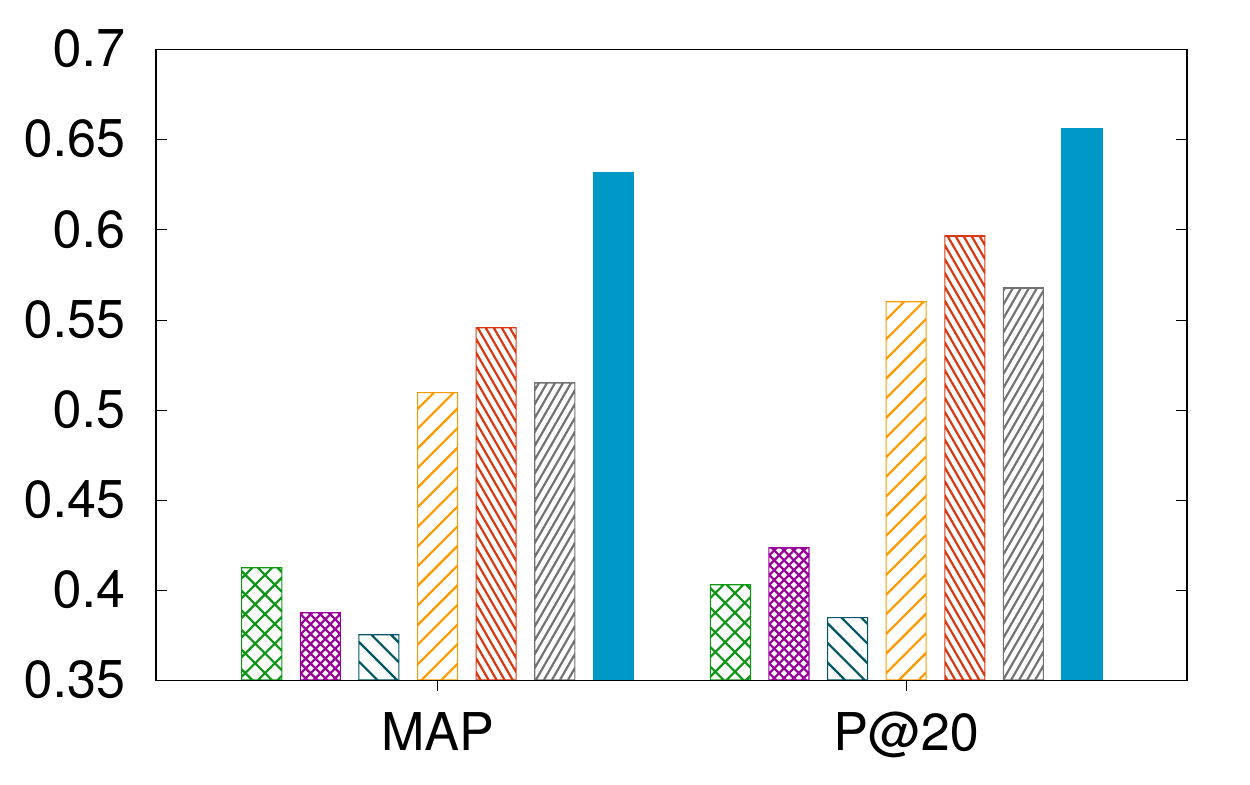}
	\caption{
		\label{fig:exp:HEPTH}
		Link prediction performance on the HepTh dataset.
		\sure shows the highest accuracies: 15.8\% higher MAP, and 10.1\% higher Precision@20 compared to the best existing method.
	}
\end{figure}

In this paper, we propose a novel relevance measure \model (RWER), an extended version of RWR, which reflects a query node's preferences on relevance scores by allowing a distinct restart probability for each node.
We also propose a supervised learning method \sure  (\underline{Su}pervised \underline{Re}start for \rwer) that automatically finds optimal restart probabilities in RWER from a given graph.
Extensive experiments show that our method provides the best the link prediction accuracy: e.g., \sure boosts MAP (Mean Average Precision) by up to 15.8\% on the best competitor as shown in Figure~\ref{fig:exp:HEPTH}.
Our main contributions are summarized as follows:
\begin{itemize*}
	\item {\textbf{Model.}
		We propose \model (\rwer), a new random walk model to improve the expressiveness of RWR. RWER allows each node to have a distinct restart probability so that the random surfer has a finer control on preferences for each node.
	}
	\item {\textbf{Learning.}
		We propose \sure, an algorithm for learning the restart probabilities in RWER from data. \sure automatically determines the best restart probabilities.
	}
	\item {\textbf{Experiment.}
		We empirically demonstrate that our proposed method improves accuracy in all dataset.
		Specifically, our proposed method improves MAP by up to 15.8\% and Precision@20 by up to 10.1\% on the best competitor. 
	}
\end{itemize*}

The code of our method and datasets used in this paper are available at \url{http://datalab.snu.ac.kr/sure}.
The rest of this paper is organized as follows.
Section~\ref{sec:prelim} presents a preliminary on RWR and defines the problem.
Our proposed methods are described in Section~\ref{sec:method}.
After presenting experimental results in Section~\ref{sec:experiment}, we provide a review on related works in Section~\ref{sec:related}.
Lastly, we conclude in Section~\ref{sec:conclusion}.

\section{Preliminaries}
\label{sec:prelim}

In this section, we describe the preliminaries on Random Walk with Restart.
Then, we formally define the problem handled in this paper.
We use $A_{ij}$ or $A(i,j)$ to denote the entry at the intersection of the $i$-th row and $j$-th column of matrix $\A$, $\A(i, :)$ to denote the $i$-th row of $\A$, and $\A(:, j)$ to denote the $j$-th column of $\A$.
The $i$-th element of the vector $\vect{x}$ is denoted by $x_i$.

\subsection{Random Walk with Restart.}
Random walk with restart (also known as Personalized PageRank, PPR)~\cite{tong2006fast} measures each node's proximity (relevance) w.r.t. a given query node $s$ in a graph.
RWR assumes a random surfer who starts at node $s$.
The surfer moves to one of its neighboring nodes with probability $1-c$ or restarts at node $s$ with probability $c$.
When the surfer moves from $u$ to one of its neighbors, each neighbor $v$ is selected with a probability proportional to the weight in the edge $(u, v)$.
The relevance score between seed node $s$ and node $u$ is the stationary probability that the surfer is at node $u$.
If the score is large, we consider that nodes $s$ and $u$ are highly related.

\begin{figure}[!t]
	\centering
	\includegraphics[width=1\linewidth]{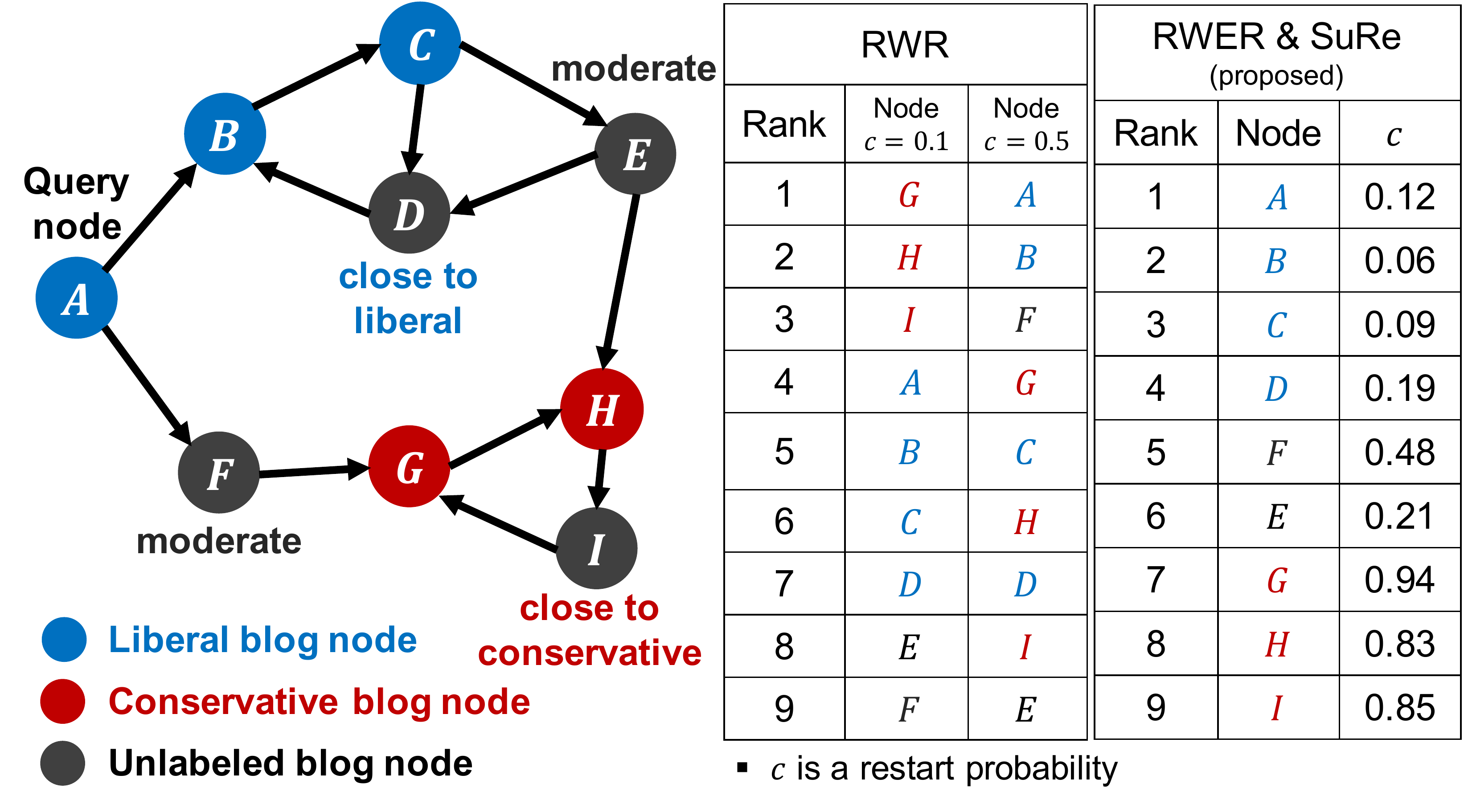}
	\caption{
		\label{fig:toy_example}
		Example of RWR and our proposed approaches \rwer \& \sure on a political blog network.
		Blue colored nodes are liberal blogs, red colored are conservative, and black colored ones are unlabeled blogs.
		RWR uses the fixed restart probability $\textbf{0.1}$ or $\textbf{0.5}$ while our proposed \rwer uses distinct restart probabilities on nodes.
		Note that RWR shows different ranking results depending on the restart probability.
		The ranking result of \rwer is more desirable for the query node $A$ (liberal) than those of RWR because many liberal blog nodes are ranked high in the ranking result.
	}
\end{figure}

\textbf{Limitations}.
RWR cannot consider a query node's preferences for estimating relevance scores between the query node and other nodes.
For example, suppose we compute relevance scores from the query node $A$ to other nodes in a political blog network in Figure~\ref{fig:toy_example} where blue colored nodes are liberal blogs, red colored ones are conservative, black ones are not labeled, and an edge between nodes indicates a hyperlink between the corresponding blogs.
Based on the topology of the graph, we consider that nodes $E$ and $F$ tend to be moderate, node $D$ is likely to be liberal, and node $I$ is likely to be conservative.
Since the query node $A$ is a liberal blog, node $A$ will prefer other liberal nodes to conservative nodes.
However, a conservative node $G$ is ranked higher than nodes related to liberal blogs such as nodes $C$ and $D$ in the ranking result of RWR.
The reason is that preferences are not considered in RWR, and
the random surfer jumps back to the query node $A$ with a fixed restart probability $c$ wherever the surfer is.
On the other hand, \rwer reflects the query node's preferences on relevance scores by allowing a distinct restart probability for each node.

Another practical problem is that it is non-trivial to set an appropriate value of the restart probability $c$ for different applications since we need to manually choose $c$ so that the restart probability provides optimal relevance scores for each application.
RWR scores are highly affected by the restart probability; the ranking results of each restart probability ($c=0.1$ and $c=0.5$) are quite different as seen in Figure~\ref{fig:toy_example}.
In contrast, our learning method \sure automatically determines the optimal restart probabilities for all nodes based on the query node's preferences as well as relationships between nodes. 
The detailed descriptions of our proposed approaches \rwer and \sure are presented in Section~\ref{sec:method}.


\subsection{Problem Definition.}
\label{pre:problem}
We are given a graph $\mathcal{G}$ with $n$ nodes and $m$ edges, a query node $s$, and side information from the query node.
The side information contains a set of \emph{positive} nodes $P = \{x_1, ..., x_k\}$ that $s$ prefers, and a set of \emph{negative} nodes $N = \{y_1, ..., y_l\}$ that $s$ dislikes.
Our task is to learn restart probabilities for all nodes such that relevance scores of the positive nodes are greater than those of the negative ones.


\section{Proposed Method}
\label{sec:method}
In this section, we describe \model (\rwer), our proposed model for extended restart probabilities. Also, we propose \sure, an efficient algorithm for learning the restart probabilities. 


\subsection{Overview of \model.}
\label{sec:method:rwer}
\rwer is a novel relevance model reflecting a query node's preferences on relevance scores.
The main idea of \rwer is that we introduce a restart probability \emph{vector} each of whose entry corresponds to a restart probability at a node, so that the restart probabilities are related to the preferences for the nodes.

In \rwer, a restart probability of each node is interpreted as the degree of boredom of a node w.r.t. the query node.
That is, if the restart probability on a node is large, then the surfer runs away from the current node to the query node (i.e., the surfer becomes bored at the node).
On the other hand, if the restart probability of the node is small, then surfer desires to move around the node's neighbors (i.e., the surfer has an interest in the node and its neighbors).

As depicted in Figure~\ref{fig:toy_example}, each node has its own restart probability in our model \rwer.
The restart probabilities are determined by our supervised learning method \sure (Section~\ref{sec:method:solution}) from the query (liberal) node $A$, the positive (liberal) nodes $B$ and $C$, and the negative (conservative) nodes $G$ and $H$.
Note that a ranking list where many liberal nodes are ranked high is desirable for the query node $A$ since $A$ is liberal.
As shown in Figure~\ref{fig:toy_example}, using distinct restart probabilities for each node by \rwer provides more satisfactory rankings for the query node than using a single restart probability for all nodes by RWR.
The restart probabilities of liberal nodes are smaller than those of conservative nodes, which implies that the random surfer prefers searching around the liberal nodes such as $B$ and $C$ while the surfer is likely to run away from the conservative nodes such as $G$ and $H$.

One might think that it is enough to simply assign small restart probabilities to positive nodes and large restart probabilities to negative nodes for a desirable ranking.
However,
the restart probabilities should be determined also for unlabeled nodes, and
the probabilities should reflect intricate relationships between nodes as well as the query node's preferences.
For example, the restart probability of node $F$ in Figure~\ref{fig:toy_example} is relatively moderate because node $F$ is located between a liberal node $A$ and a conservative node $G$.
Also, the restart probability of node $D$ is small since node $D$ is close to other liberal nodes $B$ and $C$.
Similarly, the restart probability of node $I$ is large since node $I$ is closely related to other conservative nodes $G$ and $H$.

\subsection{Formulation of \model.}
\label{sec:method:formulation}
\begin{figure}[!t]
	\centering
	\includegraphics[width=0.5\linewidth]{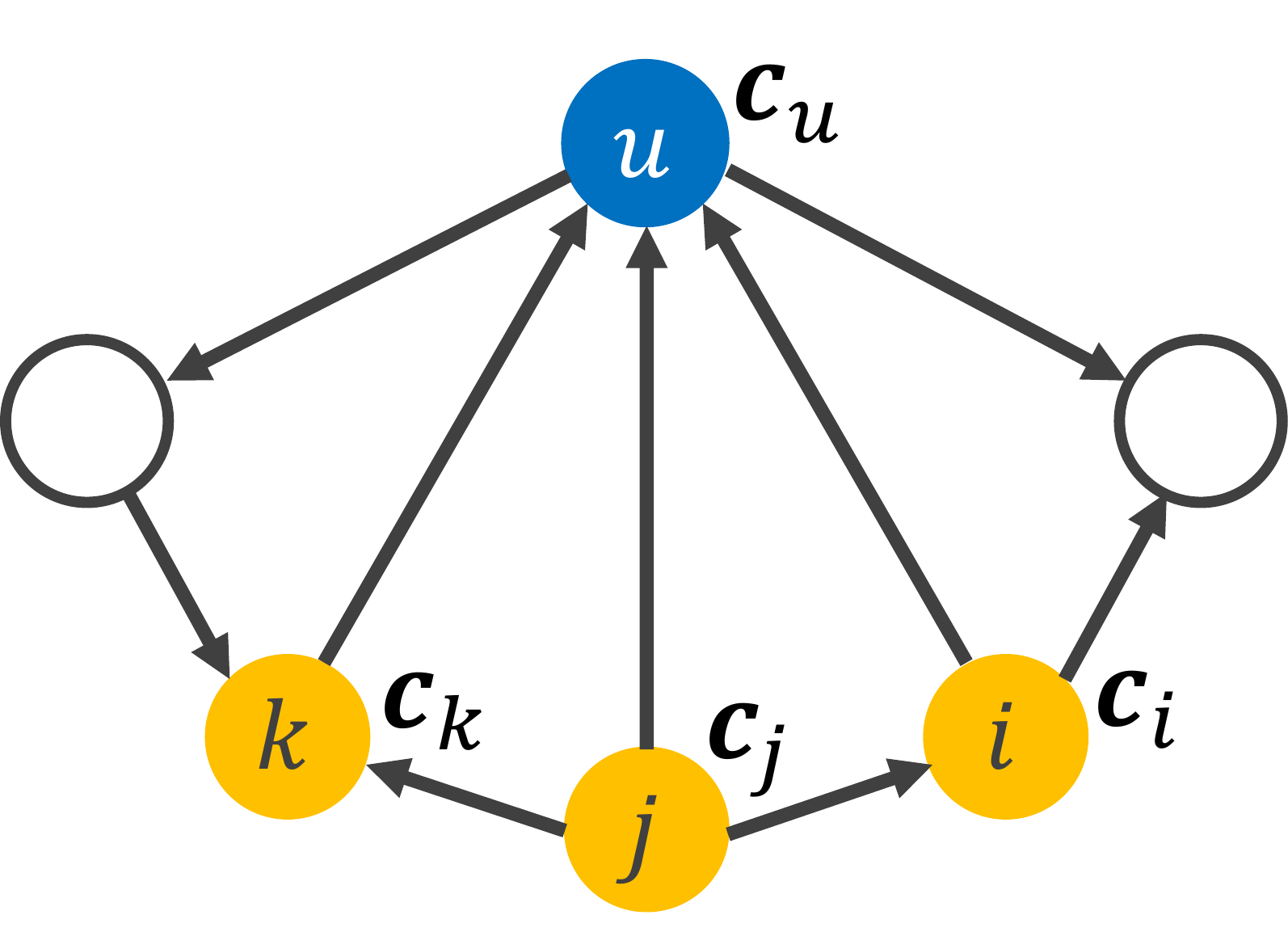}
	\caption{
		\label{fig:method:formul}
		Example of a network. Each node has its own restart probability.
	}
\end{figure}

We formulate \rwer in this section.
We first explain the formulation using the example shown in Figure~\ref{fig:method:formul}, and present general equations.
In the example, the surfer goes to one of its neighbors or jumps back to the query node.
To obtain the \rwer probability $r_u$ at time $t+1$, we should take into account the scores of the three nodes which are $i, j$ and $k$ at time $t$.
Suppose the surfer is at the node $i$ at time $t$.
The surfer can go to an out-neighbor through one of the two outgoing edges with probability $1-c_i$.
Note that every node has a distinct restart probability and node $i$ has a restart probability $c_i$ in this case.
Without the restart action, $r_u^{(t+1)}$ in Figure~\ref{fig:method:formul} is defined as follows:
\vspace{-3mm}
\small
\begin{equation*}
	\vspace{-3mm}
	r_{u}^{(t+1)} \leftarrow (1-c_i) \frac{r_{i}^{(t)}}{2} + (1-c_j) \frac{r_{j}^{(t)}}{3} + (1-c_k) r_{k}^{(t)}
\end{equation*}
\normalsize

Also, the surfer on any node $v$ jumps back to the query node with probability $c_v$.
The above equation is rewritten as follows considering the restart action of the random surfer:
\vspace{-3mm}
\small
\begin{equation*}
	\vspace{-2mm}
	\begin{split}
		r_{u}^{(t+1)} & \leftarrow (1-c_i) \frac{r_{i}^{(t)}}{2} + (1-c_j) \frac{r_{j}^{(t)}}{3} + (1-c_k) r_{k}^{(t)} \\
		& + \left(c_1 r_{1}^{(t)} + \dots+ c_v r_{v}^{(t)} + \dots + c_n r_{n}^{(t)} \right) 1(u = s)
	\end{split}
\end{equation*}
\normalsize
where $1(u = s)$ is 1 if $u$ is the query node $s$; otherwise, it is $0$.
Note that the restart term is different from that of the traditional random walk with restart.

Based on the aforementioned example, the recursive equation of our model is defined as follows:
\begin{equation}
	\label{eq:rwer_rec}
	\small
	r_{u} =  \left( \sum_{v \in \IN_{u}} (1-c_v) \frac{r_{v}}{|\OUT_{v}|} \right) + \left( \sum_{v} c_v r_v \right) 1(u = s)
	\vspace{-2mm}
\end{equation}
where $\IN_i$ is the set of in-neighbors of node $i$, and $\OUT_i$ is the set of out-neighbors of node $i$.

Equation~\eqref{eq:rwer_rec} is expressed in the form of a matrix equation as follows:
\vspace{-2mm}
\begin{equation}
	\label{eq:rwer}
	\small
	\rr=\nA^\top  (\mat{I} - diag(\cc) ) \rr + \left(\cc^\top \rr \right) \qq
	\vspace{-2mm}
\end{equation}
where $\nA$ is a row-normalized matrix of the adjacency matrix $\A$, $\cc$ is a restart vector whose $i$-th entry is $c_i$, $diag(\cc)$ is a matrix whose $diag(\cc)_{ii} = c_i$ and other entries are $0$, and $\qq$ is a vector whose $s$-th element is $1$ and all other elements are $0$.
Notice that if $\cc$ is a vector all of whose elements are the same, then the \rwer is equal to RWR (or PPR).

The following lemma shows that equation~\eqref{eq:rwer} can be represented as a closed form equation.
\begin{lemma}
	\label{lemma:closed}
	The closed form w.r.t. $\rr$ in equation~\eqref{eq:rwer} is represented as follows:
	\begin{equation}
		\small
		\rr = (\mat{I} - \mat{B})^{-1} \qq
	\end{equation}
	where $\mat{B}$ is $\nA^\top (\mat{I} - diag(\cc) ) + \qq (\cc - \mat{1})^\top $, $\nA$ is a row-normalized matrix
	, and $\vect{1}$ is an all-ones vector.
	\begin{proof}
		See Section 1.1 of~\cite{supp}.
		\QEDA
	\end{proof}
\end{lemma}

Note that if $\cc$ is given, the RWER vector $\rr$ can be calculated using the closed form in Lemma~\ref{lemma:closed}.
However, the computation using the closed form requires $O(n^3)$ time and $O(n^2)$ memory space due to the matrix inversion where $n$ is the number of nodes; thus, this approach is impractical when we need to compute RWER scores in large-scale graphs.
In order to avoid the heavy computational cost, we exploit an efficient iterative algorithm described in Section~\ref{sec:method:algo}.

\subsection{Algorithm for \model.}
\label{sec:method:algo}
We present an iterative algorithm for computing \rwer scores efficiently.
Our algorithm is based on power iteration and comprises two phases: a normalization phase (Algorithm~\ref{alg:method:normal}) and an iteration phase (Algorithm~\ref{alg:method:iter}).

\textbf{Normalization phase (Algorithm~\ref{alg:method:normal}).}
Our proposed algorithm first computes the out-degree diagonal matrix $\mat{D}$ of $\A$ (line~\ref{alg:method:normal:d}).
Then, the algorithm computes the row normalized matrix $\nA$ using $\mat{D}$ (line~\ref{alg:method:normal:nA}).

\textbf{Iteration phase (Algorithm~\ref{alg:method:iter}).}
Our algorithm computes the \rwer score vector $\rr$ for the seed node $s$ in the iteration phase.
As described in Section~\ref{sec:method:formulation}, the vector $\vect{q}$ denotes a length-$n$ starting vector whose entry at the index of the seed node is 1 and otherwise 0 (line~\ref{alg:method:iter:init}).
Our algorithm iteratively computes equation~\eqref{eq:rwer} (line~\ref{alg:method:iter:start}).
We then compute the error $\delta$ between $\rr'$, the result in the previous iteration, and $\rr$ (line~\ref{alg:method:iter:error}).
Next, we update $\rr'$ into $\rr$ for the next iteration (line~\ref{alg:method:iter:update}).
The iteration stops when the error $\delta$ is smaller than a threshold $\epsilon$ (line~\ref{alg:method:iter:stop}).

\textbf{Theoretical analysis.} We analyze the convergence of the iterative algorithm in Theorem~\ref{theorem:convergence} and the time complexity in Theorem~\ref{theorem:complexity:iter}.
We assume that all the matrices considered are saved in a sparse format, such as the compressed column storage~\cite{press1989numerical}, which stores only non-zero entries, and that all the matrix operations exploit such sparsity by only considering non-zero entries.

\begin{algorithm} [t!]
	\small
	\begin{algorithmic}[1]
		\caption{Normalization phase of \rwer} \label{alg:method:normal}
		\REQUIRE adjacency matrix $\A$
		\ENSURE row-normalized matrix $\nA$
		\STATE compute a degree diagonal matrix $\mat{D}$ of $\A$ (i.e., $\mat{D}_{ii} = \sum_{j}\A_{ij}$)\label{alg:method:normal:d}
		\STATE compute a normalized matrix, $\nA = \mati{D}\A$. \label{alg:method:normal:nA}
		\RETURN $\nA$
	\end{algorithmic}
\end{algorithm}

\begin{algorithm} [t!]
	\small
	\begin{algorithmic}[1]
		\caption{Iteration phase of \rwer} \label{alg:method:iter}
		\REQUIRE row-normalized matrix $\nA$, query node $s$, restart probability vector $\cc$, and error tolerance $\epsilon$
		\ENSURE \rwer score vector $\rr$
		\STATE set the starting vector $\qq$ from the seed node $s$ \label{alg:method:iter:init}
		\REPEAT
		\STATE $\rr' \leftarrow \nA^{\top} \left(\mat{I} - diag(\cc)\right) \rr + \left(\cc^\top \rr \right)\qq$ \label{alg:method:iter:start}
		\STATE compute error, $\delta = \lVert \rr' - \rr\rVert$ \label{alg:method:iter:error}
		\STATE update $\rr \leftarrow \rr'$ \label{alg:method:iter:update}
		\UNTIL{ $\delta < \epsilon$} \label{alg:method:iter:stop}
		\RETURN $\rr$
	\end{algorithmic}
\end{algorithm}

\begin{theorem}[Convergence]
	\label{theorem:convergence}
	Suppose the graph represented by $\nA$ is irreducible and aperiodic.
	Then, the power iteration algorithm~(Algorithm~\ref{alg:method:iter}) for \rwer converges.
	\begin{proof}
		See Section 1.2 of~\cite{supp}
	\end{proof}
\end{theorem}


\begin{theorem}[Time Complexity]
	\label{theorem:complexity:iter}
	The time complexity of Algorithm~\ref{alg:method:iter} is $O(Tm)$ where $T$ is the number of iterations, and $m$ is the number of edges.
	\begin{proof}
		See Section 1.3 of \cite{supp}.
	\end{proof}
\end{theorem}
Theorem~\ref{theorem:complexity:iter} indicates that our method in Algorithm~\ref{alg:method:iter} presents the linear scalability w.r.t the number of edges.


\subsection{Cost Function.}
\label{sec:method:problem}

Although our relevance measure \rwer improves the expressiveness of RWR by introducing a distinct restart probability for each node, it is difficult to manually investigate the optimal restart probabilities for all nodes in large graphs.
In this section, we define the cost function for finding the optimal restart probabilities.

As mentioned in Section~\ref{pre:problem}, our goal is to set the optimal restart probabilities so that the relevance scores of positive nodes outweigh those of negative nodes.
We define the following cost function:
\vspace{-2mm}
\begin{equation}
	\label{eq:general_obj}
	\small
	\argmin_{\cc}F(\cc) = \lambda \Vert \cc - \oo \Vert^2 + \sum_{x \in P, y \in N} h(r_y - r_x)
	\vspace{-2mm}
\end{equation}
where $\lambda$ is a regularization parameter that controls the importance of the regularization term, $\oo$ is a given origin vector, $h$ is a loss function, and $r_x$ and $r_y$ are \rwer scores of nodes $x$ and $y$, respectively.
The cost function is obtained from the pairwise differences between the \rwer scores of positive and negative nodes.
	Given an increasing loss function $h$, $F(\cc)$ is minimized as the scores of positive nodes are maximized and those of negative nodes are minimized.
The origin vector $\oo$ prevents the $\cc$ vector becoming too small,
and serves as a model regularizer which helps avoid overfitting and thus improves accuracy, as we will see in Section~\ref{sec:exp:parameters}.
We set $\oo$ to a constant vector all of whose elements are set to a constant.
We use the loss function $h(x) = (1+\exp(-x/b))^{-1}$ since the loss function maximizes AUC~\cite{mozer2003optimizing,backstrom2011supervised}.

\subsection{\sure \space - Optimizing the Cost Function.}
\label{sec:method:solution}
Our goal is to minimize equation~\eqref{eq:general_obj} with respect to $\cc$.
Note that the objective function $F(\vect{c})$ is not convex.
Thus, we exploit the gradient descent method to find the local minimum of function $F(\cc)$.
For the purpose, we first need to obtain the derivative of $F(\cc)$ w.r.t. $\cc$:
\vspace{-2mm}
\begin{equation}
	\label{eq:deriv_obj}
	\small
	\vspace{-2mm}
	\begin{split}
		\frac{\partial F(\cc)}{\partial \cc} & = 2 (\cc - \oo) + \sum_{x \in P, y \in N} \frac{\partial h(r_y - r_x)}{\partial \cc}\\
		& = 2 (\cc - \oo) + \sum_{x \in P, y \in N} \frac{\partial h(\delta_{yx})}{\partial \delta_{yx}} (\frac{\partial r_y}{\partial \cc} - \frac{\partial r_x}{\partial \cc})
	\end{split}
\end{equation}
where $\delta_{yx}$ is $r_y - r_x$.
The derivative $\frac{\partial h(\delta_{yx})}{\partial \delta_{yx}}$ of the loss function is $\frac{1}{b}h(\delta_{yx})(1-h(\delta_{yx}))$.

In order to obtain the derivative $\frac{\partial r_x}{\partial \cc}$, we have to calculate the derivative of the relevance score $r_x$ w.r.t. $c_i$ which is the $i$-th element of $\cc$. 
Let $\MM$ be $(\mat{I} - \mat{B})^{-1}$; then, $\rr = (\mat{I} - \mat{B})^{-1} \qq = \MM \qq$, $\MM(:,s) = \rr$, and $M(x,s)=r_x$, from Theorem~\ref{theorem:convergence}.

Since $\MM$ is the inverse of $\mat{I-B}$, according to \cite{petersen2008matrix}, $\frac{\partial \MM}{\partial c_i}$ becomes:
\vspace{-2mm}
\small
\begin{equation*}
	\frac{\partial \MM}{\partial c_i} = -\MM \frac{\partial (\mat{I} - \mat{B})}{\partial c_i} \MM = \MM (-\nA^\top \mat{J}^{ii} + \mat{J}^{si}) \MM
	\vspace{-2mm}
\end{equation*}
\normalsize
where $\mat{J}^{ij}$ is a single-entry matrix whose $(i,j)$th entry is $1$ and all other elements are $0$.
Based on the above equation, $\frac{\partial M(x,s)}{\partial c_i}$ is represented as follows:
\vspace{-2mm}
\small
\begin{equation*}
	\frac{\partial M(x,s)}{\partial c_i} = \mat{M}(x,:) (-\nA^\top(:, i) + \mat{e}_s) M(i,s)
	\vspace{-2mm}
\end{equation*}
\normalsize
where $\mat{e}_s$ is a length $n$ unit vector whose $s$-th entry is 1.
Note that $\frac{\partial M(x,s)}{\partial c_i}$ is calculated for $1\leq i \leq n$; then, $\frac{\partial r_x}{\partial \cc}$ is written in the following equation:
	\begin{equation}
		\label{eq:deriv_r}
		\small
			\frac{\partial r_x}{\partial \cc} = \frac{\partial M(x,s)}{\partial \cc} = \left((-\nA + \mat{1} \mat{e}_s^\top) \MM(x,:)^\top \right) \circ \MM(:,s)
\end{equation}
\noindent where $\circ$ denotes Hadamard product, and $\mat{1}$ is an all-ones vector of length $n$.
Similarly,
$\frac{\partial r_y}{\partial \cc}$ is calculated by switching $x$ to $y$.

	Using the equation~\eqref{eq:deriv_r}, $\sum_{x\in P, y\in N}\frac{\partial h(\delta_{yx})}{\partial \delta_{yx}} (\frac{\partial r_y}{\partial \cc} - \frac{\partial r_x}{\partial \cc} )$ in equation \eqref{eq:deriv_obj} is represented as follows:
	\begin{align}
	\small
		\begin{split}
		\label{eq:drdc}
		\left( (-\nA + \mat{1} \mat{e}_{s}^{\top}) \!\!\!\! \sum_{x \in P, y \in N} \!\!\! \frac{\partial h(\delta_{yx})}{\partial \delta_{yx}}
		\left( \MM(y,:)\!-\!\MM(x,:) \right)^{\top} \right) &\circ \MM(:,s) \\
		= \left( (-\nA + \mat{1} \mat{e}_{s}^{\top}) \tilde{\rr}  \right) &\circ \rr
		\end{split}
		\vspace{-3mm}
	\end{align}
	\noindent where $\rr = \MM(:,s)$ is an \rwer score vector, and $\tilde{\rr} = \sum_{x \in P, y \in N} \!\! \frac{\partial h(\delta_{yx})}{\partial \delta_{yx}} \left( \MM(y,:)-\MM(x,:)\right)^\top$.
	Then, $\frac{\partial F(\cc)}{\partial \cc}$ in equation~\eqref{eq:deriv_obj} is represented as follows:
\begin{equation}
\small
\label{eq:dfdc}
\frac{\partial F(\cc)}{\partial \cc} =  2 (\cc - \oo) + \left( (-\nA + \mat{1} \mat{e}_{s}^{\top}) \tilde{\rr} \right) \circ \rr
\end{equation}

\begin{algorithm} [t!]
	\small
		\begin{algorithmic}[1]
			\caption{\sure \space- Learning a restart vector $\cc$} \label{alg:learning:restart}
			\REQUIRE adjacency matrix $\A$, query node $s$, positive set $P$, negative set $N$, origin vector $\oo$, parameter $b$ of loss function $h$, and the learning rate $\eta$
			\ENSURE the learned restart vector $\cc$
			\STATE initialize $\cc \leftarrow \oo$
			\WHILE {$\cc$ does not converge}
			\STATE compute $\rr = \MM(:,s)$ based on equation~\eqref{eq:rwer}  (Algorithm~\ref{alg:method:iter})
			\STATE compute $\tilde{\rr} = \sum_{x \in P, y \in N} \frac{\partial h(\delta_{yx})}{\partial \delta_{yx}}	\left( \MM(y,:)-\MM(x,:)\right)^\top$ by a linear system solver (Lemma~\ref{lemma:method:Mx}) 
			\STATE compute $\frac{\partial F(\cc)}{\partial \cc}$ by equation~\eqref{eq:dfdc} \label{alg:learning:restart:obj}
			\STATE update $\cc \leftarrow \cc - \eta \frac{\partial F(\cc)}{\partial \cc}$ \label{alg:learning:restart:update}
			\ENDWHILE
			\RETURN the learned restart vector $\cc$
		\end{algorithmic}
\end{algorithm}

Notice that we do not obtain $\MM$ explicitly to compute $\MM(:,s)$ in equations~\eqref{eq:dfdc} since $\MM$ is the inverse of $\mat{I-B}$ and inverting a large matrix is infeasible as mentioned in Section~\ref{sec:method:formulation}.
Instead, we use the iterative method described in Algorithm~\ref{alg:method:iter} to get $\rr=\MM(:,s)$.
However, the problem is that we also require rows of $\MM$ (i.e., $\MM(x, :)$ in $\tilde{\rr}$), and Algorithm~\ref{alg:method:iter} only computes a column of $\MM$ for a given seed node.
How can we calculate $\tilde{\rr}$ without inverting $\MM$?
$\tilde{\rr}$ is computed iteratively by the following lemma:



\begin{lemma}
	\label{lemma:method:Mx}
	From the result of equation~\eqref{eq:drdc}, $\tilde{\rr}=\sum_{x \in P, y \in N} \frac{\partial h(\delta_{yx})}{\partial \delta_{yx}}	\left( \MM(y,:) \!-\MM(x,:)\right)^\top$ which is represented as $\tilde{\rr}=\mat{M}^{\top}\tilde{\vect{p}} \Leftrightarrow (\mat{I} - \mat{B}^{\top})\tilde{\rr} = \tilde{\vect{p}}$ where $\tilde{\vect{p}} = \sum_{x \in P, y \in N} \!\! \frac{\partial h(\delta_{yx})}{\partial \delta_{yx}} (\vect{e}_y - \vect{e}_x)$, $\mat{e}_x$ is an $n \times 1$ vector whose $x$-th element is $1$ and the others are $0$, and $ \delta_{yx} = r_y - r_x$.
	Then, $\tilde{\rr}$ is the solution of the linear system $(\mat{I} - \mat{B}^{\top})\tilde{\rr} = \tilde{\vect{p}}$ which is solved by an iterative method for linear systems.
	\begin{proof}
		See Section 1.4 of~\cite{supp}.
		\QEDA
	\end{proof}
\end{lemma}

Note that $\MM^{-\top}=\vect{I} - \vect{B}^\top$ is non-symmetric and invertible (Lemma 1.1. of \cite{supp}); thus, any iterative method for a non-symmetric matrix can be used to solve for $\tilde{\rr}$.
	We use GMRES~\cite{saad1986gmres}, an iterative method for solving linear systems since it is the state-of-the-art method in terms of efficiency and accuracy.

\begin{figure}[t!]
	\centering
	\hspace{-4mm}
	\includegraphics[width=1\linewidth]{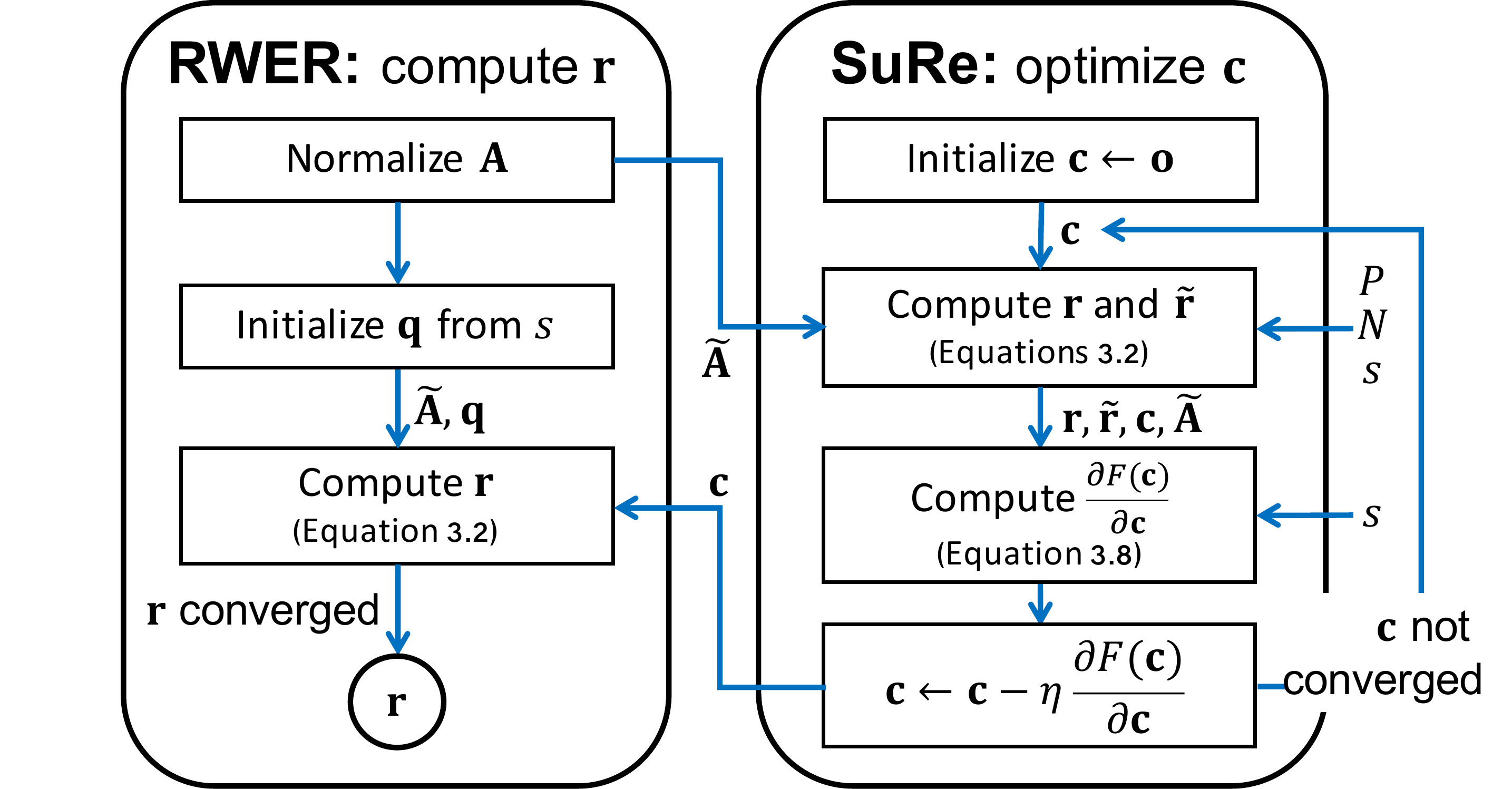}
	\caption{Flowchart of \rwer (Algorithms~\ref{alg:method:normal} and \ref{alg:method:iter}) and \sure (Algorithm~\ref{alg:learning:restart}).
			\sure learns restart probability vector $\cc$, and \rwer computes our node relevance score vector $\rr$ for given seed node $s$. }
	\label{fig:method:flowchart}
\end{figure}

\textbf{Optimization phase (Algorithm~\ref{alg:learning:restart}).}
\sure algorithm for solving the optimization problem is summarized in Algorithm~\ref{alg:learning:restart} and Figure~\ref{fig:method:flowchart}.
In the algorithm, we use a gradient-based method to update the restart probability $\cc$ based on equation~\eqref{eq:dfdc}.

\subsection{Theoretical analysis.}
\label{sec:method:analysis}
We analyze the time complexity of \sure (Algorithm~\ref{alg:learning:restart}).


\begin{lemma}
	\label{lemma:method:derivtime}
	Let $|P|$ and $|N|$ denote the number of positive and negative nodes, respectively.
		The computation of $\tilde{\rr}=\sum_{x, y} \frac{\partial h(\delta_{yx})}{\partial \delta_{yx}}(\MM(y,:)-\MM(x,:))^\top$ takes $O(Tm+|P||N|)$ time where $T$ is the number of iterations until convergence, and $m$ is the number of edges.
		\begin{proof}
			See Section 1.5 of \cite{supp}.
			\QEDA
		\end{proof}
\end{lemma}

Based on Lemma~\ref{lemma:method:derivtime}, the time complexity of Algorithm~\ref{alg:learning:restart} is presented in Theorem~\ref{theorem:complexity:time}.
\begin{theorem}[Time complexity of Algorithm~\ref{alg:learning:restart}]
	\label{theorem:complexity:time}
		For a given graph with $m$ non-zero elements,
		the learning algorithm \sure takes $O(T_1 (T_2 m + |P||N|)) $ time where $T_1$ is the number of times $\cc$ is updated with the gradient, and $T_2$ is the number of inner iterations for computing $\rr$ and $\tilde{\vect{\rr}}$.
		\begin{proof}
			See Section 1.6 of \cite{supp}.
			\QEDA
		\end{proof}
\end{theorem}

Theorem~\ref{theorem:complexity:time} implies that our learning method \sure in Algorithm~\ref{alg:learning:restart} provides linear time and space scalability w.r.t. the number $m$ of edges.
Notice that $|P|$ and $|N|$ are constants much smaller than $m$.

\section{Experiment}
\label{sec:experiment}
We evaluate our proposed method \sure with various baseline approaches.
Since there is no ground-truth of node-to-node relevance scores in real-world graphs,
we instead evaluate the performance of two representative applications based on relevance scores: ranking and link prediction.
Based on these settings, we aim to answer the following questions from the experiments:

\begin{itemize*}
	\item{\textbf{Q1.} \textbf{Ranking performance (Section~\ref{sec:exp:ranking}).}
		Does our proposed method \sure provide the best relevances scores for ranking compared to other methods?}
	\item{\textbf{Q2.} \textbf{Link prediction performance (Section~\ref{sec:exp:link_pred_perfor}).} How effective is \sure for link prediction tasks?}
	\item{\textbf{Q3.} \textbf{Parameter sensitivity (Section~\ref{sec:exp:parameters}).} How does the value of the origin vector used in \sure affect the accuracy of link prediction?}
	\item{\textbf{Q4.} \textbf{Scalability (Section~\ref{sec:exp:scalability}).} How well does \sure scale up with the number of edges?}
\end{itemize*}


\begin{table}[!t]
	\centering
	\small
	\caption{Dataset statistics.
		The query nodes are used for the ranking and the link prediction tasks.
	}
	\begin{threeparttable}[t]
		\begin{tabular}{c|rrrl}
			\toprule
			\textbf{Dataset} & \textbf{\#Nodes} & \textbf{\#Edges}  & \textbf{\#Queries}\\
			\midrule
			Wikipedia\tnote{1} & 3,023,165 & 102,382,410 & - \\
			Epinions\tnote{1} & 131,828 & 841,327 & 200\\
			Slashdot\tnote{1} & 79,120 & 515,397 & 200\\
			HepPh\tnote{1} & 34,546 & 421,534   & 135 \\
			HepTh\tnote{1} & 27,770 & 352,768   & 121 \\
			Polblogs\tnote{2} & 1,490 & 19,025  & 115 \\
			\bottomrule
		\end{tabular}
		\begin{tablenotes}

			\item[1] {\url{http://konect.uni-koblenz.de}}
			\item[2] {\url{http://www-personal.umich.edu/~mejn/netdata}}
		\end{tablenotes}
	\end{threeparttable}
	\label{tab:dataset}
\end{table}

\subsection{Experimental Settings.}
\hfill

\textbf{Datasets.}
We experiment on various real-world network datasets.
Datasets used in our experiments are summarized in Table~\ref{tab:dataset}.
We use Polblogs and signed networks (Epinions and Slashdot) for the ranking task (Section~\ref{sec:exp:ranking}), HepPh and HepTh for the link prediction task (Section~\ref{sec:exp:link_pred_perfor}), and Wikipedia for the scalability experiment (Section~\ref{sec:exp:scalability}).
Since only HepPh and HepTh have time information, we use them in the link prediction task.
All experiments are performed on a Linux machine with Intel(R) Xeon E5-2630 v4 CPU @ 2.2GHz and 256GB memory.

\textbf{Methods.}
We compare our proposed method \sure with
Common Neighbor (CN)~\cite{liben2007link}, Adamic-Adar (AA)~\cite{adamic2003friends}, Jaccard's Coefficient (JC)~\cite{salton1986introduction}, Random Walk with Restart (RWR)~\cite{haveliwala2002topic}, MRWR~\cite{shahriari2014ranking}, Supervised Random Walks (SRW)~\cite{backstrom2011supervised}, and QUINT~\cite{li2016quint}.
We set parameters in each method to the ones that give the best performance (see Section 2 of \cite{supp} for parameters).

\textbf{Evaluation Metrics.}
To compare the methods, we use Mean Average Precision (MAP), Area under the ROC curve (AUC), and Precision$@20$.
MAP is the mean of average precisions for multiple queries.
AUC is the expectation that a uniformly drawn random positive is ranked higher than a uniformly drawn random negative.
Precision$@20$ is the precision at the top-$20$ position in a ranking result.
The higher the values of the metrics are, the better the performance is.

\begin{table}[t!]
	\centering
	\caption{
		Ranking results of our proposed method \sure and other methods w.r.t. a query node \textit{obsidianwings}, a liberal blog.
		Red colored nodes are conservative blogs, and the black colored ones are liberal blogs.
		Our ranking result from \sure contains only liberal nodes, indicating the best result, while other ranking results wrongly contain conservative nodes.
	}
	\scriptsize
	\label{tab:exp:polblogs:1hop}
	\begin{tabular}{c|c|c|c|c}
		\toprule
		\textbf{Rank} & \textbf{\sure} & \textbf{RWR} & \textbf{SRW} & \textbf{QUINT} \\
		\midrule
		1  & digbysb & \redtext{\textbf{freerep}} & digbysb & \redtext{\textbf{freerep}} \\
		2  & billmon    & \redtext{\textbf{michell}} & tbogg      & \redtext{\textbf{michell}} \\
		3  & gadflye   & \redtext{\textbf{prolife}} & liberal & \redtext{\textbf{prolife}} \\
		4  & reachm     & \redtext{\textbf{rightwi}} & billmon    & \redtext{\textbf{rightwi}} \\
		5  & jameswo & digbysb & xnerg      & digbysb \\
		6  & angrybe  & \redtext{\textbf{littleg}} & corrent   & \redtext{\textbf{littleg}} \\
		7  & marksch & billmon    & \redtext{\textbf{hughhew}} & billmon    \\
		8  & tbogg      & jameswo & busybus & jameswo \\
		9  & wampum     & reachm     & pacific & reachm     \\
		10 & stevegi & politic & nielsen & \redtext{\textbf{hughhew}} \\
		\bottomrule
	\end{tabular}
	\label{tab:ranking}
\end{table}

\begin{figure}[t!]
	\vspace{3mm}
	\centerline
	{\includegraphics[width=0.9\linewidth]{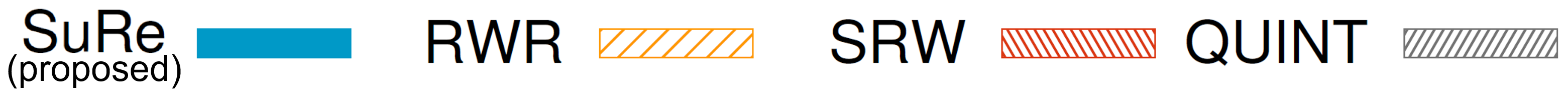}}
	\subfigure[MAP]
	{
		\label{fig:exp:pol:map}
		\includegraphics[width=0.467\linewidth]{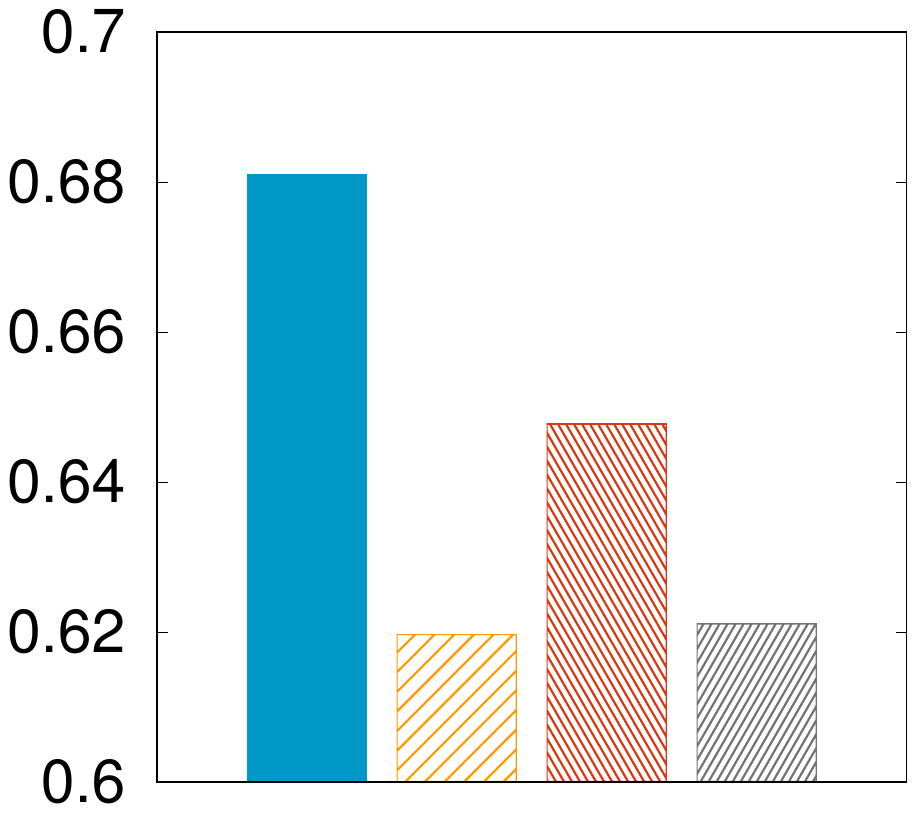}
	}
	\subfigure[Precision@20]
	{
		\label{fig:exp:pol:pre}
		\includegraphics[width=0.467\linewidth]{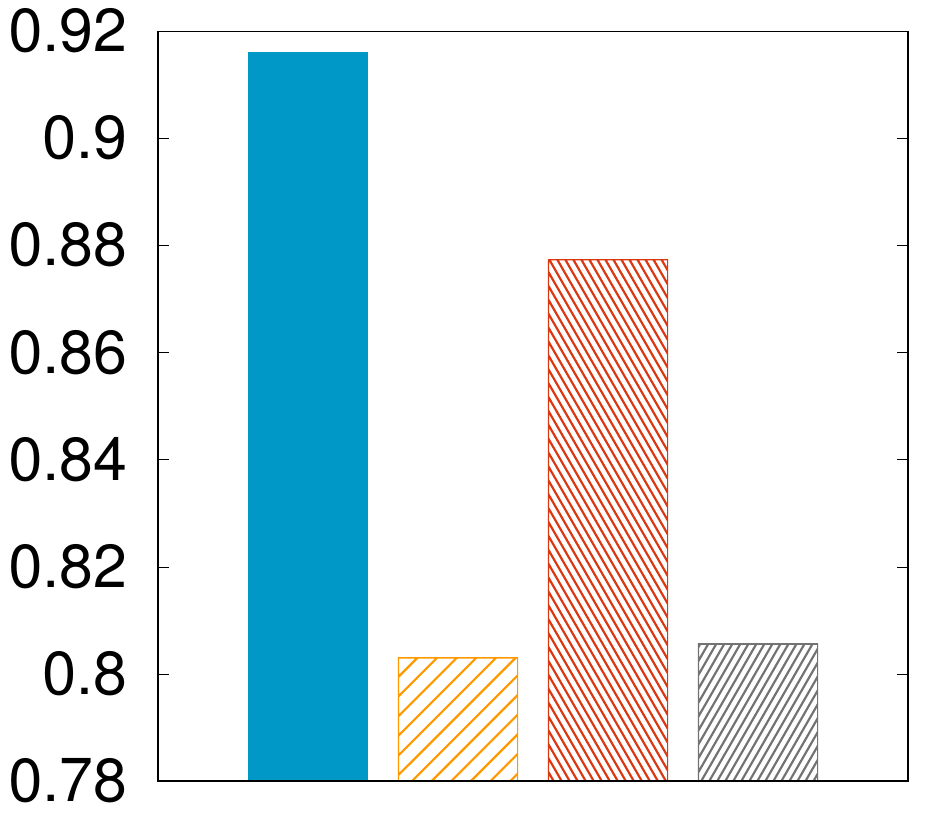}
	}
	\caption{
		\label{fig:exp:pol}
		Ranking performance on Polblogs. Our method \sure provides the best ranking performance compared to other methods in terms of MAP and Precision@20.
	}
\end{figure}

\subsection{Ranking Performance.}
\label{sec:exp:ranking}

We evaluate the ranking performance of our method \sure compared to that of other methods.

\textbf{Experimental setup.}
We perform this experiments on the Polblogs dataset and signed networks (Epinions and Slashdot).
See Section 2.1 of~\cite{supp} for detailed experimental setup.

%

\textbf{Case study.}
We analyze the ranking quality produced from each method in the Polblogs dataset.
Table~\ref{tab:ranking} shows the top-10 ranking list for a query node \textit{obsidianwings}, a liberal blog.
Red colored nodes are conservative, and the black colored ones are liberal.
As shown in the table, our ranking result from \sure is of a higher quality compared to those from RWR, SRW, and QUINT since top-10 ranking result from \sure contains only liberal nodes while other ranking results have several conservative nodes, considering that the query node is liberal.

\begin{figure}[t!]
	\centerline{\includegraphics[width=0.65\linewidth]{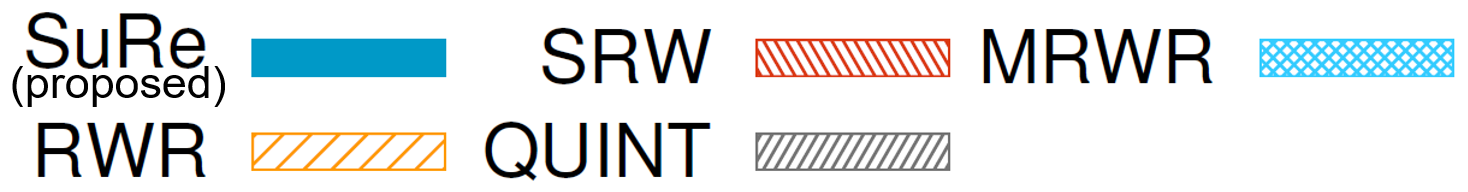}}
	\subfigure[Epinions]
	{
		\label{fig:exp:sign:epinions}
		\includegraphics[width=0.467\linewidth]{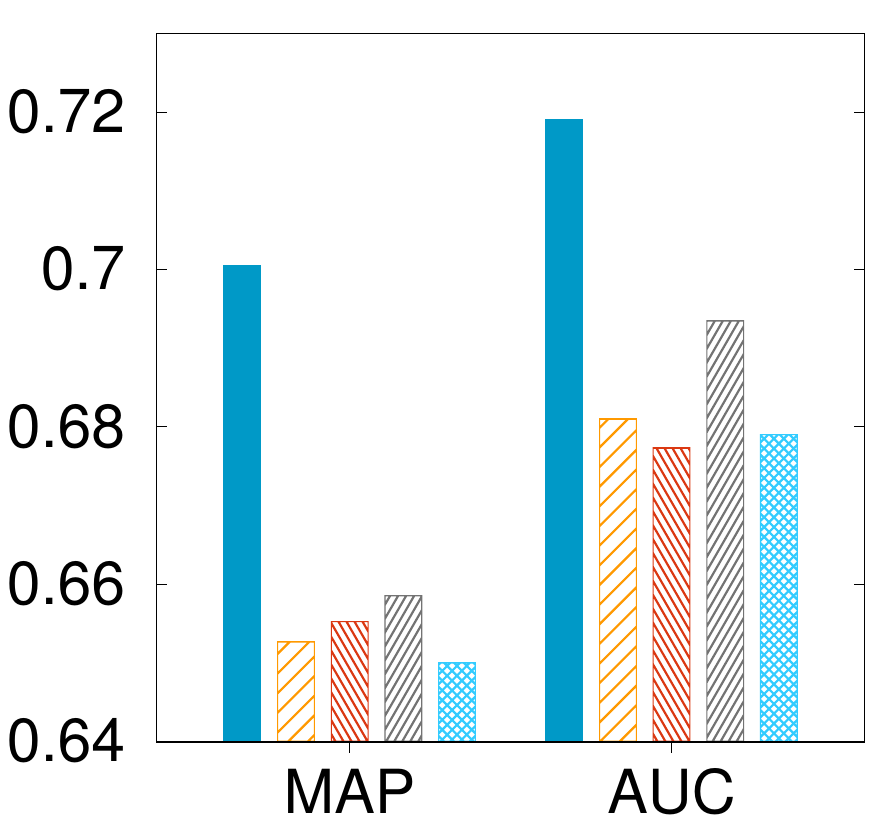}
	}
	\subfigure[Slashdot]
	{
		\label{fig:exp:sign:slashdot}
		\includegraphics[width=0.467\linewidth]{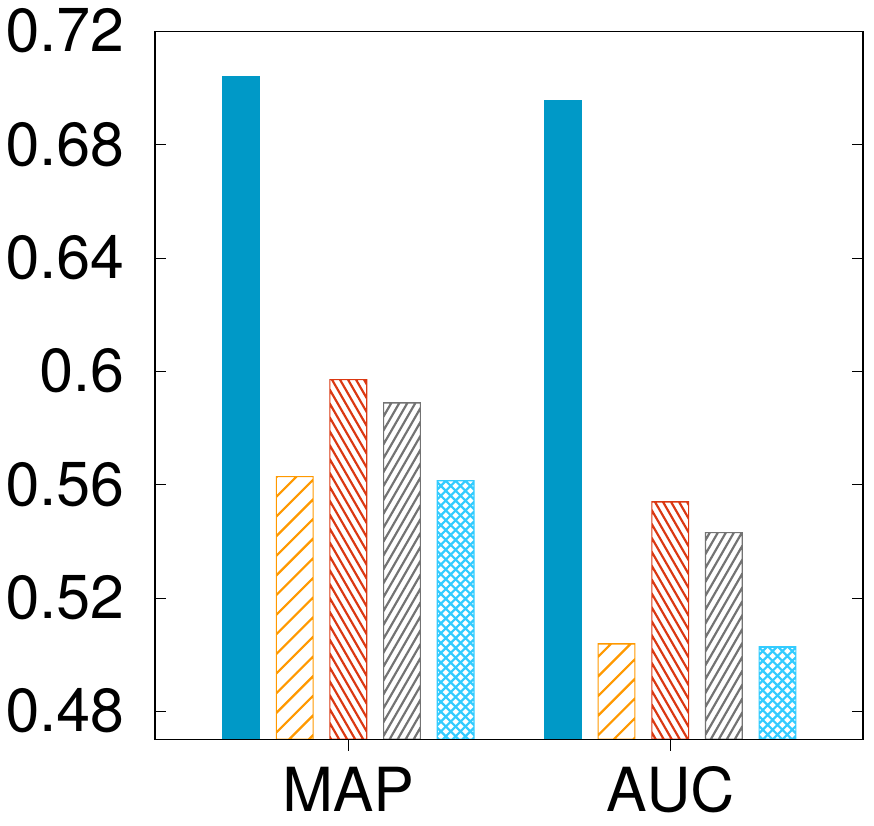}
	}
	\caption{
		\label{fig:exp:sign}
		Ranking performance on the signed networks.
		Our method \sure obtains up to 17.9\% higher MAP and 25.6\% higher Precision@20 (in Slashdot) compared to the best competitor.
	}
\end{figure}

\begin{figure}[t!]
	\label{fig:exp}
	\centering
	{\includegraphics[width=0.85\linewidth]{FIG/bar/labels3.png}}
	\includegraphics[width=0.8\linewidth]{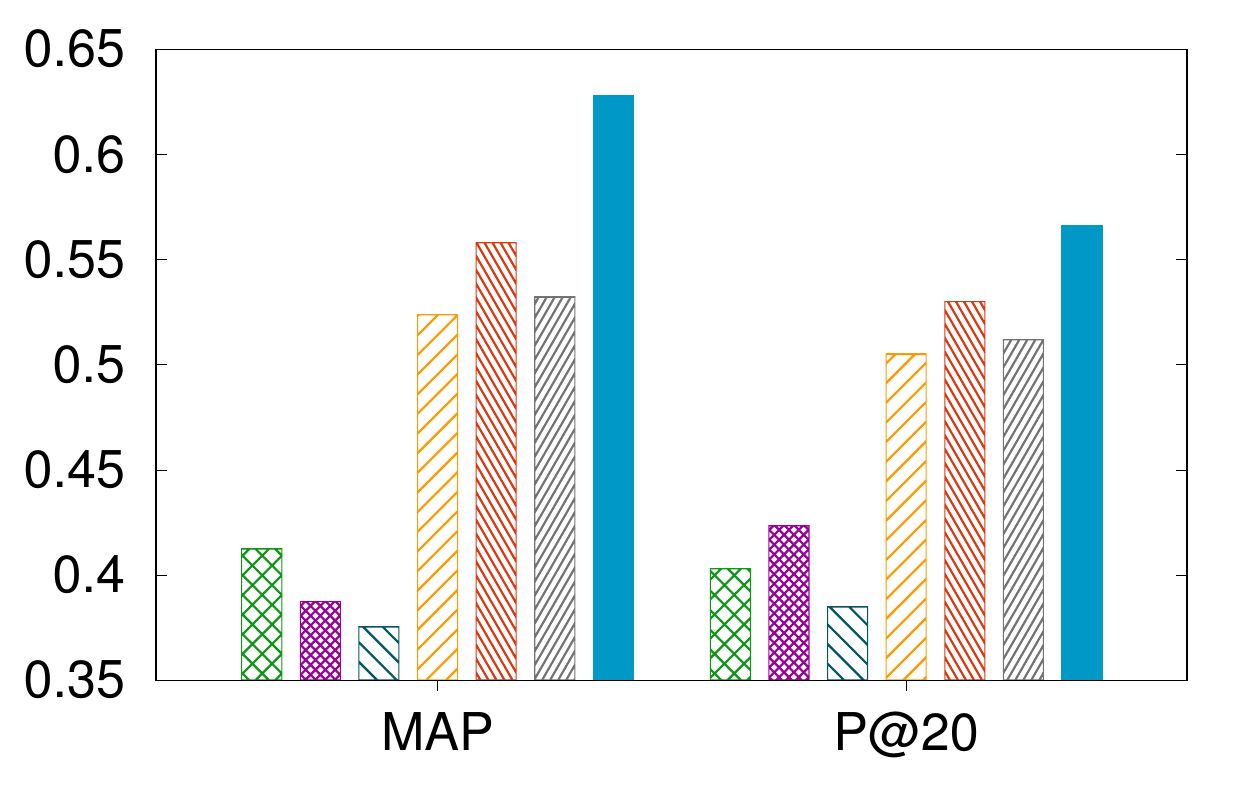}
	\caption{
		\label{fig:exp:HEPPH}
		Link prediction performance on the HepPh dataset.
		\sure shows the highest accuracies: 12.5\% higher MAP, and 6.8\% higher Precision@20 compared to the best existing method.
	}
\end{figure}

\textbf{Result.}
To evaluate ranking performances, we measure MAP, Precision@20, and AUC for the ranking results produced from our method \sure including other random walk based methods.
For brevity, we report MAP and Precision@20 in Polblogs, and MAP and AUC in Epinions and Slashdot.
	
	In Polblogs, if the query node is liberal (conservative), then the positive class is liberal (conservative), and the negative one is conservative (liberal).
	As shown in Figure~\ref{fig:exp:pol}, our method \sure shows the best ranking performance compared to other methods in terms of MAP and Precision@20.

In signed networks, we evaluate the performance of \sure compared to other baselines including MRWR~\cite{shahriari2014ranking}, an RWR-based method for signed networks.
	As in Figure~\ref{fig:exp:sign}, our method \sure shows the best performance: up to 17.9\% higher MAP and 25.6\% higher Precision@20 compared to the best competitor.

\subsection{Link Prediction Performance.}
\label{sec:exp:link_pred_perfor}

We examine the link prediction performance of our proposed method \sure compared to other link prediction methods as well as RWR-based methods SRW and QUINT.


\textbf{Experimental Setup.}
We perform this experiments on the HepPh and HepTh datasets which are time-stamped networks.
	See Section 2.2 of~\cite{supp} for detailed experimental setup.

\textbf{Result.} Figures~\ref{fig:exp:HEPTH} and \ref{fig:exp:HEPPH} show the link prediction performances in terms of MAP and Precision@$20$.
As shown in the results, our method \sure outperforms other competitors including SRW and QUINT which are the state-of-the-art methods for link prediction.
%
In the HepTh dataset, compared to the best competitor SRW, \sure achieves $15.8\%$ improvement in terms of MAP, and $10.1\%$ improvement in terms of Precision@20 (Figure~\ref{fig:exp:HEPTH}).
For the AUC results, see Section 2.3 of \cite{supp}.
Note that \sure provides the best prediction over all datasets.
The results state that assigning a distinct restart probability to each node and learning the restart probabilities (\rwer and \sure) have a significant effect on link prediction compared to using a fixed restart probability for all nodes (RWR).
Furthermore, the result indicates that learning restart probabilities (\sure) provides better link prediction accuracy than existing supervised learning methods that focus on learning edge weights (SRW) or network topology (QUINT).

\begin{table}[!h]
	\centering
	\small
	\caption{Number of parameters for each method.}
	\begin{tabular}{c|ccc}
		\toprule
		Dataset & SRW & \textbf{\sure}  & QUINT\\
		\midrule
		$\# parameters$ &    $O(\# features)$   & $O(n)$	    & $O(n^2)$ \\
		\bottomrule
	\end{tabular}
	\label{tab:result:parameters}
\end{table}

\textbf{Discussion.}
	We discuss the above experimental results in terms of the number of model parameters.
	As shown in Table~\ref{tab:result:parameters}, SRW has not enough parameters (i.e., $\# features < n$); thus, feature selection is important for the performance of applications in SRW.
	On the other hand, QUINT has too many parameters; thus, QUINT is prone to overfit.
	Also, it is infeasible to learn $O(n^2)$ parameters in large-scale graphs.
	Compared to these methods, \sure has a moderate number of parameters, implying that 1) the expressiveness of \sure is better than that of SRW, and 2) \sure is not likely to overfit compared to QUINT.
	This point explains why \sure provides better performance of the ranking and link prediction tasks than SRW and QUINT do as shown in Sections~\ref{sec:exp:ranking}~and~\ref{sec:exp:link_pred_perfor}.

\subsection{Parameter Sensitivity.}
\label{sec:exp:parameters}
\begin{figure}[t!]
	\centering
	\hspace{-3mm}
	\subfigure[MAP of ranking task in the Polblogs dataset]
	{
		\label{fig:exp:restart:map:polblogs}
		\includegraphics[width=0.477\linewidth]{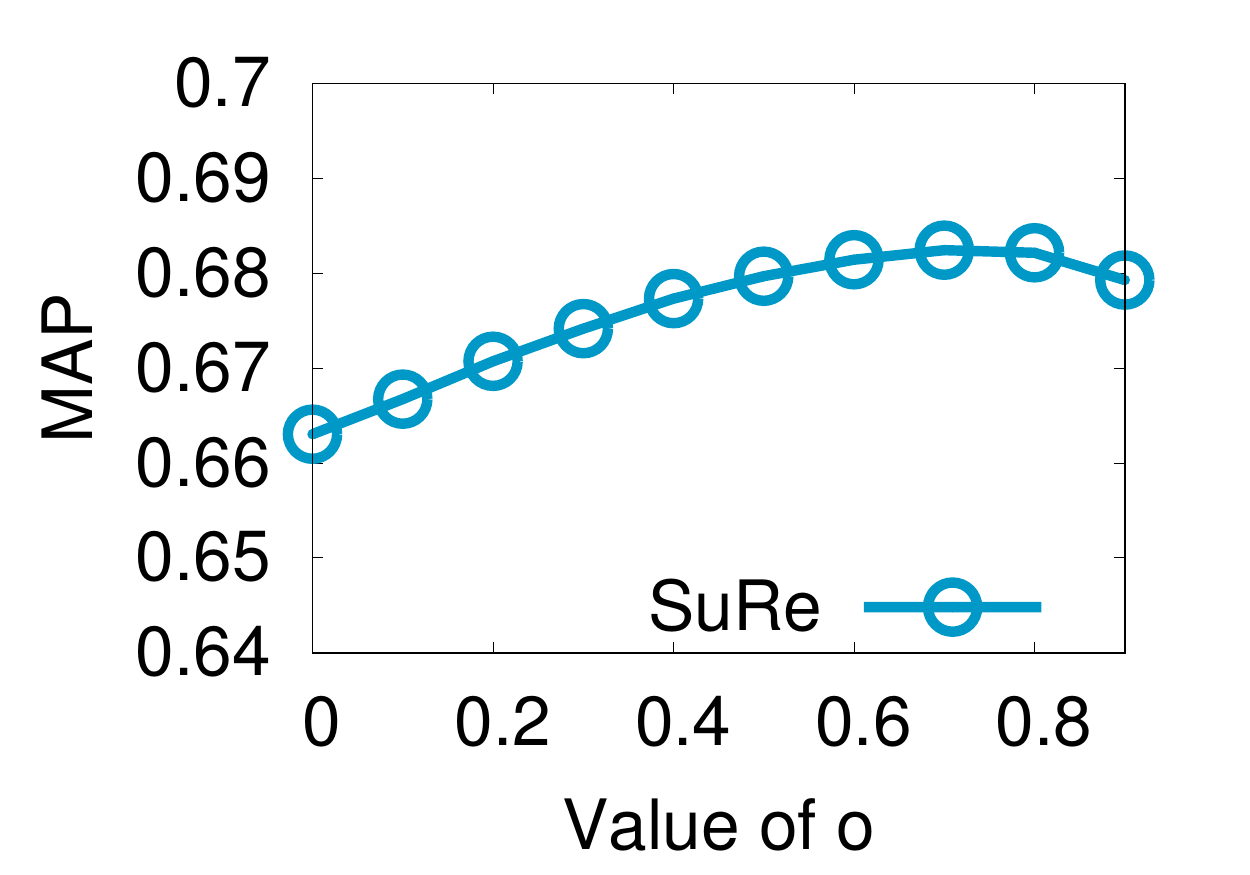}
	}
	\subfigure[MAP of link prediction task in the HepTh dataset]
	{
				\label{fig:exp:restart:map:cora}
		\includegraphics[width=0.477\linewidth]{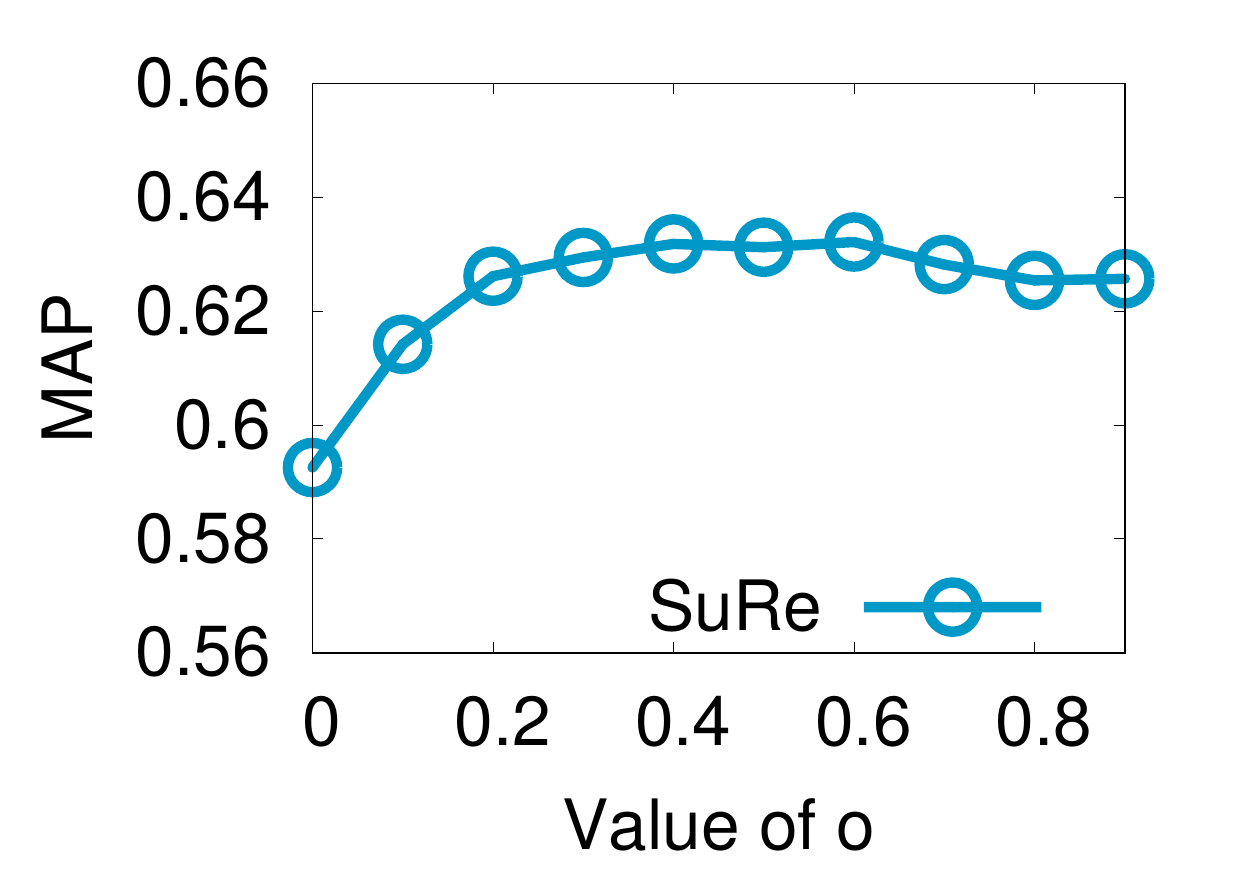}
	}
	\caption{
		\label{fig:exp:restart}
		Parameter sensitivity of our method \sure in Polblogs and HepTh datasets.
		We report the link prediction and ranking accuracy using MAP measure,
			changing the values of the elements in the origin vector $\vect{o}$ in \sure, where all the elements of $\vect{o}$ is set to a same value.
Note that the performance of \sure is improved by introducing the origin parameter $\oo$.
	}
\end{figure}

We investigate the parameter sensitivity of \sure w.r.t. the value of the origin vector $\oo$.
The origin vector $\oo$ serves as a model regularizer
which helps avoid overfitting and improves accuracy, as described in Section~\ref{sec:method:problem}.
We evaluate MAP of ranking and link prediction tasks, and report the results in Figure~\ref{fig:exp:restart}.
Note that the performance of \sure is improved by introducing the origin parameter $\oo$, compared to the case without $\oo$, which corresponds to the leftmost points in both plots of Figure~\ref{fig:exp:restart}.

\subsection{Scalability.}
\label{sec:exp:scalability}
\begin{figure}[t!]
	\centering
	{
		 \includegraphics[width=0.7\linewidth]{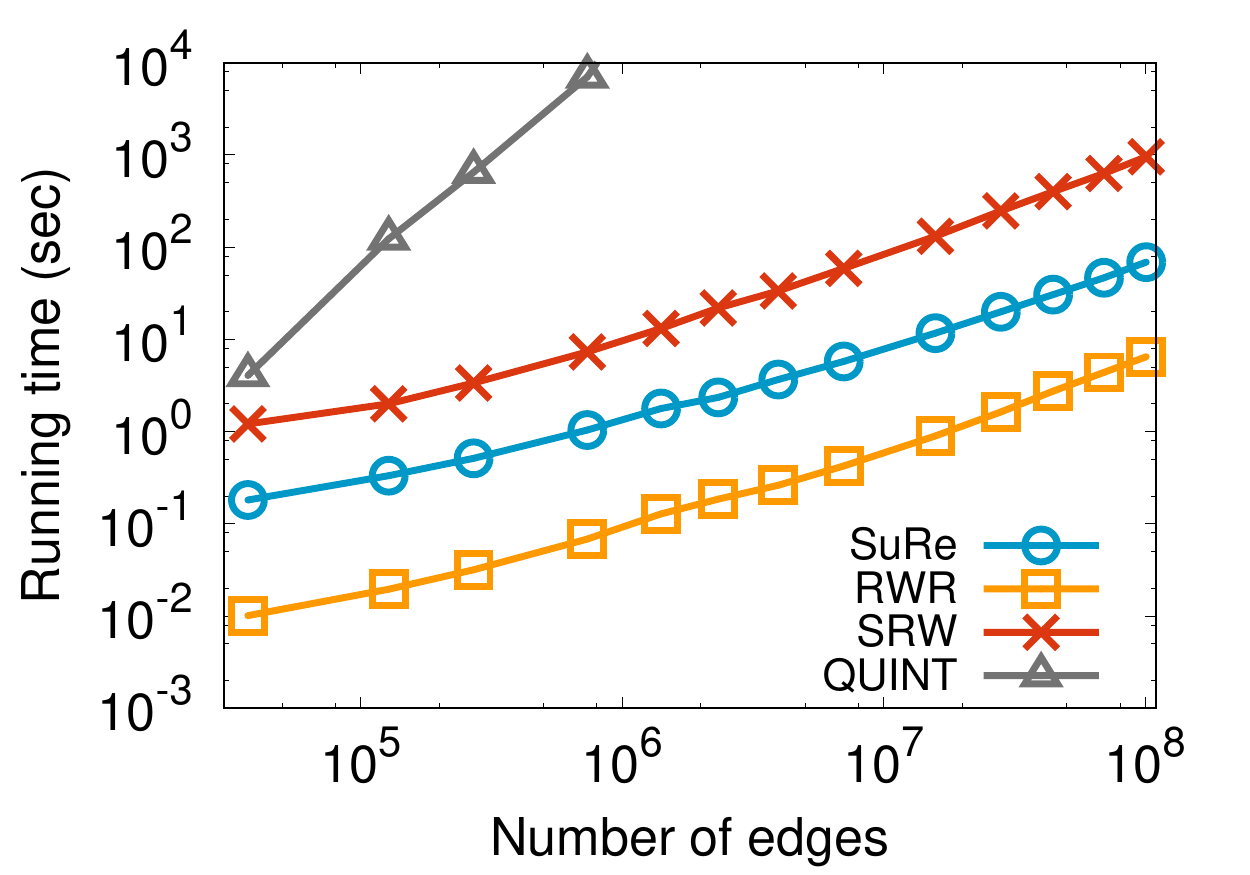}
	}
	\caption{
		\label{fig:exp:scale}
		Scalability of \sure with other baselines in the Wikipedia dataset. 
			The figure shows that \sure has near-linear scalability w.r.t. the number of edges.
	}
\end{figure}

We examine the scalability of our proposed method \sure compared to other baselines.
We perform the optimization phase in Algorithm~\ref{alg:learning:restart} with various sizes of the Wikipedia dataset to investigate the scalability of \sure.
	Figure~\ref{fig:exp:scale} shows that \sure scales near-linearly with the number of edges. 
	The slope of the fitted line for \sure is $0.76$, the smallest number: those for RWR, SRW, and QUINT are $0.83$, $0.88$, and $2.57$, respectively.
	As shown in Figure~\ref{fig:exp:scale}, \sure is not the fastest among the tested methods, but \sure is the fastest among all learning based methods (SRW and QUINT).
	Note that the result for \sure is consistent with Theorem~\ref{theorem:complexity:time} which states that \sure scales near-linearly w.r.t. the number of edges.

\section{Related Works}
\label{sec:related}

The related works fall into two main categories: 1) relevance measures in graphs, and 2) ranking and link prediction based on relevance measures.


\textbf{Relevance measures in graphs.}
There are various relevance measures in graphs based on link analysis and random walk, e.g., PageRank~\cite{page1999pagerank}, HITS~\cite{kleinberg1999authoritative}, Random Walk Graph Kernel~\cite{kang2012fast}, and RWR (or Personalized PageRank)~\cite{haveliwala2002topic}.
Among these measures, RWR has received much attention from the data mining community since it provides a personalized ranking w.r.t. a node, and it has been applied to many graph mining applications such as community detection~\cite{andersen2006local}, link prediction~\cite{backstrom2011supervised,li2016quint}, ranking~\cite{tong2006fast}, and graph matching~\cite{tong2007fast}.
Also, fast and scalable methods~\cite{shin2015bear,jung2016random,tong2006fast} for computing RWR in large graphs have been proposed to boost the performance of those applications in terms of time.

\textbf{Ranking and link prediction.}
Jung et al.~\cite{srwr_ICDM2016} extended the concept of RWR to design a personalized ranking model in signed networks.
Wang et al.~\cite{wang2006image} proposed an image annotation technique that generates candidate annotations and re-ranks
them using RWR.
Liben-Nowell et al.~\cite{liben2007link} extensively studied the link prediction problem in social networks based on relevance measures such as PageRank, RWR, and Adamic-Adar~\cite{adamic2003friends}.
%
Many researchers have proposed supervised learning methods for link prediction.
Backstrom et al.~\cite{backstrom2011supervised} proposed Supervised Random Walk (SRW), a supervised learning method for link prediction based on RWR.
SRW learns parameters for adjusting edge weights.
%
Li et al.~\cite{li2016quint} developed QUINT, a learning method for finding a query-specific optimal network.
QUINT modifies the network topology including edge weights.
In many real-world scenarios, however, modifying the graph structure would not be allowed.
On the contrary, our \sure method controls the behavior of the random surfer without modifying the graph structure, and provides better prediction accuracy than other competitors as shown in Section~\ref{sec:experiment}.

\section{Conclusion}
\label{sec:conclusion}
We propose \model (\rwer), a novel relevance measure using distinct restart probabilities for each node.
We also propose \sure, a data-driven algorithm for learning restart probabilities of \rwer.
Experiments show that our method brings the best performance for ranking and link prediction tasks, outperforming the traditional RWR and recent supervised learning methods.
Specifically, \sure improves MAP by up to 15.8\% on the best competitor. 
Future works include 
designing distributed algorithms for computing and learning \rwer.

\bibliography{BIB/myref}
\bibliographystyle{abbrv}


\end{document}